# Towards high-performance photodetectors based on quasi-2D Ruddlesden-Popper mixed-n perovskite nanomaterials

Brindhu Malani S,[a] Eugen Klein,[a] Rostyslav Lesyuk [ab] and Christian Klinke[*acd]

[a] *Institute of Physics, University of Rostock, Albert-Einstein-Straße 23, 18059 Rostock, Germany*

[b] *Pidstryhach Institute for Applied Problems of Mechanics and Mathematics of NAS of Ukraine, Naukowa Str. 3b, 79060 Lviv, Ukraine*

[c] *Department Life, Light & Matter, University of Rostock, Albert-Einstein-Strasse 25, 18059 Rostock, Germany Department of Chemistry, Swansea University, Swansea, United Kingdom*

[d] *Department of Chemistry, Swansea University – Singleton Park, Swansea SA2 8PP, United Kingdom*

[*] *Corresponding Author: Christian.klinke@uni–rostock.de*

**Abstract**

The excellent optoelectronic properties, straightforward synthesis, and robust material stability of quasi-2D metal halide perovskites have made them a hot research topic for optoelectronic devices. In layered organic-inorganic perovskites, the optoelectronic properties are greatly influenced by morphology and quantum confinement, which are crucial for their photodetection performance and offer an excellent tunability tool. Here we investigated the optoelectronic properties and photodetection performance of quasi-2D methylammonium lead bromide perovskites with three different morphologies nanoplatelets, nanosheets, and nanostripes produced by colloidal hot injection technique. The structural and optical characterization reveals the mixed-n phases with varying extents of confinement. Our study shows that nanostripes with the weak confinement effect exhibit superior photodetection performance. They demonstrate highest photoresponsivity, EQE, and detectivity of 183 mA W$^{-1}$, 56 % and 2.9 x 10$^{11}$ Jones respectively. They also show the highest on/off ratio and fastest photoresponse ($\tau_r$=3.4 ms and $\tau_f$=4.3 ms) in comparison to nanoplatelets and nanosheets. Our study will aid in realizing an inexpensive high-performance photodetector based on nanoparticles of quasi-2D perovskites.

# Introduction

Metal halide perovskites possess exceptional optoelectronic properties, rendering them highly suitable for numerous optoelectronic applications including photodetection. These materials offer higher absorption coefficients, high carrier mobility, long-range carrier diffusion, high photoluminescence quantum yield, and tunable bandgap energies.[1,2] Bulk (3D) perovskites have an $ABX_3$ crystal structure, where A sites are cations ($Cs^+$, $FA^+$ and $MA^+$), B sites are also cations ($Pb^{2+}$, $Sn^{2+}$, $Bi^{3+}$ and $Ge^{2+}$) and X are halides ($Cl^-$ and $I^-$).[2–4] Despite their exceptional optoelectronic properties, their commercial usage is challenging due to their low long-term chemical stability in moist environments.[5] The single crystalline perovskite exhibits a higher absorption coefficient than crystalline Si, but it is difficult to realize a photodetector with a thick macroscopic size.[6] However, polycrystalline 3D perovskites suffer from high defect densities, low carrier mobilities and higher charge recombination due to grain boundaries.[7] The low-dimensional perovskites possess similar optoelectronic properties and can be produced easily on a large scale on flexible substrates.[8] In recent years, organic-inorganic hybrid 2D layered perovskites have generated wide interest due to their enhanced stability in ambient conditions. The most common is the Ruddlesden-Popper (RP) crystal structure $P`_2P_{n-1}Q_nX_{3n+1}$, where P` is a long chain cation, P is a smaller organic cation, Q is a metal cation, X is halide anion and n is the number of QX layers between neighboring spacers (n=1 to ∞), when n approaches infinity the structure goes to bulk (3D) perovskite.[9,10] The $QX_6$ blocks are separated by an alkyl-long chain forming multiple quantum wells. The motion of charge carriers is restricted in these planes resulting in quantum and dielectric confinement.[11,12] The optoelectronic properties are tunable with n, as n decreases, confinement increases, leading to a rise in binding energy much higher than in their 3D counterparts.[9–15] The higher binding energy and insulating hydrophobic ligands limit the generation and transport of free carriers at room temperature and affect photodetection applications.[15] Increasing the layer number n can reduce binding energy due to a weakening of dielectric confinement. The binding energy and confinement effect also depend on the size of perovskite particles. The quasi-2D perovskite materials are intermediate between 2D and 3D possessing lower trap density, better carrier mobility, carrier diffusion, and stability.[16,17] There has been extensive research on photodetectors focusing on 3D perovskites while comparatively fewer on low dimensions.[12,18] The influence of morphology and quantum confinement on photodetection in quasi-2D perovskite is not well understood.[13–15]

Mostly, active layers of photodetector are based on traditional crystalline inorganic semiconductors such as Si, Ge, II-VI, and III-V compounds.[19] These photodetectors have been extensively investigated and have a complex fabrication process requiring high temperature and expensive instrumentations such as molecular beam epitaxy, thermal deposition, and chemical vapor deposition. In contrast, hybrid 2D layered perovskite photodetectors are fabricated by simple solution deposition of colloidal perovskites synthesized by the hot-injection method.[10,12,20,21] These methods are cost-effective and suitable for mass production.

In this work, we have investigated the photodetection performance of quasi-2D $MAPbBr_3$ perovskites with different morphologies as nanoplatelets, nanosheets, and nanostripes with varying extents of quantum confinement. The absorption and PL studies reveal that the nanostripes exhibit weaker confinement with predominant high-n phases. In contrast, nanoplatelets and nanosheets have stronger confinement with predominant low-n phases. All three morphologies of quasi-2D $MAPbBr_3$ perovskite show low dark current due to confinement effect, surface ligands and charge carrier hopping between neighboring particles. The nanostripes exhibit the highest photoresponsivity, EQE and detectivity of 183 mA $W^{-1}$, 56 % and 2.9 x $10^{11}$ Jones respectively. They also demonstrate the fastest photoresponse with a rise time of 3.4 ms and a fall time of 4.3 ms. The binding energy and confinement effects increase with size reduction. Our research shows that the nanostripes due to their morphology (which bridges the electrode) and weaker confinement demonstrate better photodetection performance. The nanostripes have mixed properties of bulk and confinement which gives the advantage of both. We believe that our study will help optimize the quantum confinement effect for developing stable and high-performance photodetectors using quasi-2D perovskites.

## Experimental section

### Synthesis of nanoplatelets, nanostripes, and nanosheets of MAPbBr$_3$

**Chemicals and reagents:** All chemicals were used as received: Lead(II) acetate tri-hydrate (Aldrich, 99.999%), nonanoic acid (Alfa Aesar, 97%), tri-octylphosphine (TOP; ABCR, 97%), 1-bromotetradecan (BTD; Aldrich, 97%), methylammonium bromide (MAB; Aldrich, 98%), diphenyl ether (DPE; Aldrich, 99%), toluene (VWR, 99,5%), dimethylformamide (DMF; Aldrich, 99,8%), dodecylamine (DDA; Merck, 98%), acetone (HPLC grade, LiChrosolv, Merck, Germany).

**PbBr$_2$ nanosheet synthesis:** In a typical synthesis a three-neck 50 mL flask was used with a condenser, septum and thermocouple. 860 mg of lead acetate tri-hydrate (2.3 mmol) were dissolved in 10 mL of nonanoic acid (57 mmol) and 10 mL of 1-bromotetradecane (34 mmol) and heated to 75 °C until the solution turned clear in a nitrogen atmosphere. Then vacuum was applied to remove the acetic acid which is generated by the reaction of nonanoic acid with the acetate from the lead precursor. After 1.5 h the reaction apparatus was filled with nitrogen again and the reaction was started by adding 0.06 mL of TOP (0.13 mmol) at a temperature of 140 °C and was stopped 8 minutes later. After 8 minutes the heat source was removed and the solution was left to cool down below 80 °C. Afterward, it was centrifuged one time at 4000 rpm for 3 minutes. The particles were suspended in 7.5 mL toluene and put into a freezer for storage.

**Synthesis of MAPbBr$_3$ nanoplatelets:** A three-neck 50 mL flask was used with a condenser, septum and thermocouple. 10.5 mL of diphenyl ether (66.1 mmol), 0.2 mL of a 500 mg dodecylamine (2.69 mmol) in 4 mL diphenyl ether precursor) were heated to 80 °C in a nitrogen atmosphere. At 80 °C 0.1 mL of trinoctylphosphine (0.22 mmol) was added. Then vacuum was applied to dry the solution. After 1.5 h the reaction apparatus was filled with nitrogen again and the temperature was set to 100 °C. 1 mL of as-prepared PbBr$_2$ nanosheets in toluene was added and heated until everything was dissolved. The synthesis was started with the injection of 0.02 mL of a 300 mg methylammonium bromide (2.68 mmol) in 6 mL dimethylformamide precursor. After 2 minutes the heat source was removed and the solution was left to cool down below 60 °C. Afterward, it was centrifuged at 4000 rpm for 3 minutes. The particles were washed two times in toluene before the product was finally suspended in toluene again.

**Synthesis of MAPbBr$_3$ nanostripes:** A three-neck 50 mL flask was used with a condenser, septum and thermocouple. 6 mL of diphenyl ether (37.7 mmol), 0.08 mL of a 500 mg dodecylamine (2.69 mmol) in 4 mL diphenyl ether precursor were heated to 80 °C in a nitrogen atmosphere. At 80 °C 0.2 mL of trinoctylphosphine (0.45 mmol) was added. Then vacuum was applied to dry the solution. After 1.5 h the reaction apparatus was filled with nitrogen again and the temperature was set to 130 °C. 0.8 mL of as-prepared PbBr$_2$ nanosheets in toluene were added and heated until everything was dissolved. The synthesis was started with the injection of 0.05 mL of a 300 mg methylammonium bromide (2.68 mmol) in 6 mL dimethylformamide precursor. After 10 minutes the heat source was removed and the solution was left to cool down below 60 °C. Afterward, it was centrifuged at 4000 rpm for 3 minutes. The particles were washed two times in toluene before the product was finally suspended in toluene again.

**Synthesis of MAPbBr$_3$ nanosheets:** A three-neck 50 mL flask was used with a condenser, septum and thermocouple. 10.5 mL of diphenyl ether (66.1 mmol), 0.2 mL of a 500 mg dodecylamine (2.69 mmol) in 4 mL diphenyl ether precursor) were heated to 80 °C in a nitrogen atmosphere. At 80 °C 0.2 mL of trinoctylphosphine (0.45 mmol) was added. Then vacuum was applied to dry the solution. After 1.5 h the reaction apparatus was filled with nitrogen again and the temperature was set to 160 °C. 2.5 mL of as-prepared $PbBr_2$ nanosheets in toluene were added and heated until everything was dissolved. The synthesis was started with the injection of 0.06 mL of a 300 mg methylammonium bromide (2.68 mmol) in 6 mL dimethylformamide precursor. The heating continued until all initially formed perovskite material completely dissolved. Then, the heat source was removed and the solution was left to cool down below 60 °C. The crystallization started at 110 °C. Afterward, it was centrifuged at 4000 rpm for 3 minutes. The particles were washed two times in toluene before the product was finally suspended in toluene again.

**Device fabrication and measurement of photodetector:** The $Si/SiO_2$ (300 nm ± 5% thermal oxide, resistivity of .005-.020 Ω-cm, Addison Engineering, Inc. San Jose CA) substrate with interdigitated gold electrodes with an electrode spacing of 5 μm was used for fabricating photodetectors. These substrates were washed with acetone for 2 min in an ultrasonic bath, followed by drying with nitrogen gas. The 40 μL solution of colloidal $MAPbBr_3$ nanoparticles in toluene is drop-casted on interdigitated gold electrodes, then annealed at 50°C for 2 min in a vacuum (Lakeshore 340 temperature controller) for photodetection measurement. The active area of each device is estimated using SEM and fluorescence microscopic images. The Keithley 4200 semiconductor characterization system was used for characterizing the I-V curves of the device under different illumination power densities and bias voltages. The laser with an excitation wavelength of $\lambda_{ex}$ = 405 nm was chosen to excite all the n-phases from n=2 to n=∞ simultaneously. The transient current response of the device is measured under different bias voltages and power densities. Using a function generator (RIGOL, DG4062), the laser light was turned into pulsed light to measure the response time. The current frequency response is measured under a bias voltage of 4 V (SRS70, power supply) and power density of 5.2 mW/cm$^2$ with pulsed laser light and an oscilloscope (Tektronix, TDS2014B).

**Characterization:** The TEM samples were prepared by diluting the nanostripe suspension with toluene followed by drop casting 10 μL of the suspension on a TEM copper grid coated with a carbon film. Standard images were done on a Talos-L120C and EM-912 Omega with a thermal emitter operated at an acceleration voltage of 120 kV and 100 kV.

UV/vis absorption spectra were obtained with a Lambda 1050+ spectrophotometer from Perkin Elmer equipped with an integration-sphere. The photoluminescence (PL) spectra measurements were obtained by a fluorescence spectrometer (Spectrofluorometer FS5, Edinburgh Instruments). For the time-resolved photoluminescence (TRPL) measurements, a picosecond laser with 375 nm excitation wavelength and 100 kHz repetition rate was used. The decay profiles are tail-fitted with a bi-exponential function $R(t) = A_1 \exp\left(-\frac{t}{\tau_1}\right) + A_2 \exp\left(-\frac{t}{\tau_2}\right)$ and the average PL lifetime is calculated using the formula $\tau_{average} = \frac{A_1\tau_1^2 + A_2\tau_2^2}{A_1\tau_1 + A_2\tau_2}$.

Photoluminescence quantum yield (PLQY) of the samples were measured using an absolute method by directly exciting the sample solution and the reference (toluene in our case) in an SC-30 integrating sphere module fitted to a Spectrofluorometer FS5 from Edinburg Instrument. During the measurement, the excitation slit was set to 6.5 nm, and the emission slit was adjusted to obtain a signal level of 1×10$^6$ cps. A wavelength step size of 0.1 nm and an integration time of 0.2 s were used. The calculation of absolute PL QY is based on the formula, $\eta = \frac{E_{sample} - E_{ref}}{S_{ref} - S_{sample}}$, where $\eta$ is absolute PL QY, $E_{sample}$ and $E_{ref}$ are the integrals at the emission region for the sample and the reference, respectively, and $S_{sample}$ and $S_{ref}$ are the integrals at the excitation scatter region for the sample and the reference, respectively. The selection and calculation of integrals from the emission and excitation scattering region and the calculation of absolute PLQY were performed using the FLUORACLE software from the Edinburg Instrument.

A scanning electron microscope (SEM, Zeiss EVO/MA10) was used for imaging of deposited colloidal MAPbBr$_3$ nanoplatelets, nanostripes and nanosheets on interdigitated gold electrodes. We used the Inbeam secondary electron mode at an operating voltage of 10 kV, a working distance of 10 mm, and a magnification of ×370.

Fluorescence images were obtained from a fluorescence microscope (Amscope) with an excitation wavelength of 400 nm using an objective of 20X and NA of 0.40. The fluorescence and SEM images were analyzed using ImageJ software to calculate the active area for each device.

An atomic force microscope (AFM, Park Systems XE-100) in non-contact mode (ARROW-NCR-20, force constant $k = 42\,\text{Nm}^{-1}$, resonance frequency at 285 kHz) was used to obtain surface topography.

**Result and Discussion**

The bright field TEM images of three different morphologies, nanoplatelets, nanostripes, and nanosheets of quasi-2D MAPbBr$_3$ synthesized through the colloidal hot injection method with slightly different reaction parameters (precursor concentration, temperature, and time) are shown in Figure 1.[10] The nucleation and growth processes as well as the parameter-dependent changes in the size and shape of 2D MAPbBr$_3$ nanostructures were described in the previous article.[22] As depicted in Figure 1 (a) the formed nanoplatelets exhibit shapes that vary between square and rectangle. There is a wide size distribution from 40 nm to 190 nm. Nanostripes are rectangular with lengths ranging from 2.2 μm to 8 μm and widths ranging from 25 nm to 290 nm (Figure 1 (c)). Nanosheets have similar shapes as nanoplatelets, appearing as rectangles and squares, but they are larger, ranging from 3.5 μm to 6 μm (Figure 1 (e)). The selected area electron diffraction (SAED) of quasi-2D MAPbBr$_3$ is shown in Figure S1. All three morphologies display dot patterns indicating the presence of crystallinity. The AFM images of the quasi-2D MAPbBr$_3$ are shown along with the height profiles in Figure S2. The nanoplatelets show the smallest thickness, ranging from 2 to 13 nm, while nanostripes have a thickness range of 6 to 37 nm. The nanosheets show a maximum thickness ranging from 25 to 232 nm due to their highly stacked layered structure.

Figure 1 (b, d, f) presents the UV-visible absorbance and PL measurements for these nanocrystals. Nanoplatelets show strong absorbance peaks at 2.99 eV, 2.79 eV, 2.65 eV, 2.56 eV and a weaker feature at 2.42 eV, which is attributed to different phases with *n* = 2, 5, 7, 8 and ∞ (referred to as bulk phase), respectively.[23] Similarly, nanostripes and nanosheets show multiple absorbance peaks corresponding to different *n* phases as listed in Table S1. The nanostripes' spectra significantly differ from the nanoplatelets and nanosheets. The former exhibits stronger spectral features related to high-n phases (n>8, bulk-like) and the latter demonstrates excitonic features related to low-n phases (n≤8). Both nanoplatelets and nanosheets do not show bulk phase (n=∞) in the absorption spectra, which is exclusively observed in the nanostripes. However, nanostripes display significant spectral features for n=3, along with dominant bulk phases. This indicates that nanostripes are mostly composed of high-n phases with a small fraction of low-n phases. In contrast, nanoplatelets and nanosheets primarily consist of low-n phases, with only a small fraction being high-n phases. Nanosheets exhibit strong spectral features for low-n phases (n=5, 3), and a weak spectral feature for n=8 and 7, suggesting that they predominantly consist of low-n phases. The nanoplatelets exhibit strong spectral features for n=7, 5, and 2 with a weak spectral feature for n=8, indicating that their n-phases are intermediate between those of nanostripes and nanosheets. The nanostripe absorbance edge is redshifted revealing a reduced band gap energy and exciton binding energy compared to the nanoplatelets and nanosheets.[13] In terms of dielectric and quantum confinement, nanosheets exhibit the strongest effect, followed by nanoplatelets and then nanostripes.

The normalized PL spectra exhibit a strong emission around 2.40 eV and a weak emission at higher energies originating from the confinement effect due to low-n phases. The absorbance and PL results confirm the spontaneous formation of mixed-n phases in nanoparticles with varying compositions of high-n and low-n phases. The TRPL recorded for these nanoparticles are presented in Figure S3 with its fitting parameters listed in Table S2. The nanoplatelets and nanosheets have a longer lifetime compared to nanostripes. In nanoplatelets and nanosheets, the TRPL photophysics is governed by majority excitons, which funnel from low-n phases into high-n phases.[24,25] The isolated small fractions of high-n phases in them act as traps, which collect the excitons through funneling. As reported in an earlier study on these nanoparticles funneling (10-50 ps) can surpass the self-trapping resulting in longer lifetimes.[23] The nanoplatelets show the highest PLQY of 51%. This indicates the potential for improved photodetection performance. However, the effective transport of generated carriers to electrodes critically depends on the morphology, spatial distribution and the relative amount of high-n phases in nanoparticles. The X-ray powder diffraction (XRD) characterization (Figure S4) confirms the mixing of n-phases and bulk phase in our samples especially well pronounced in nanosheets and nanostripes. In each XRD diffractogram, one dominant n-phase in combination with bulk phase can be clearly recognized as being complementary to the steady-state UV-visible absorbance and PL spectroscopy. Latter serves as the main source of information about the mixing of n-phases and provides completeness of the picture due to larger sensitivity in this case.

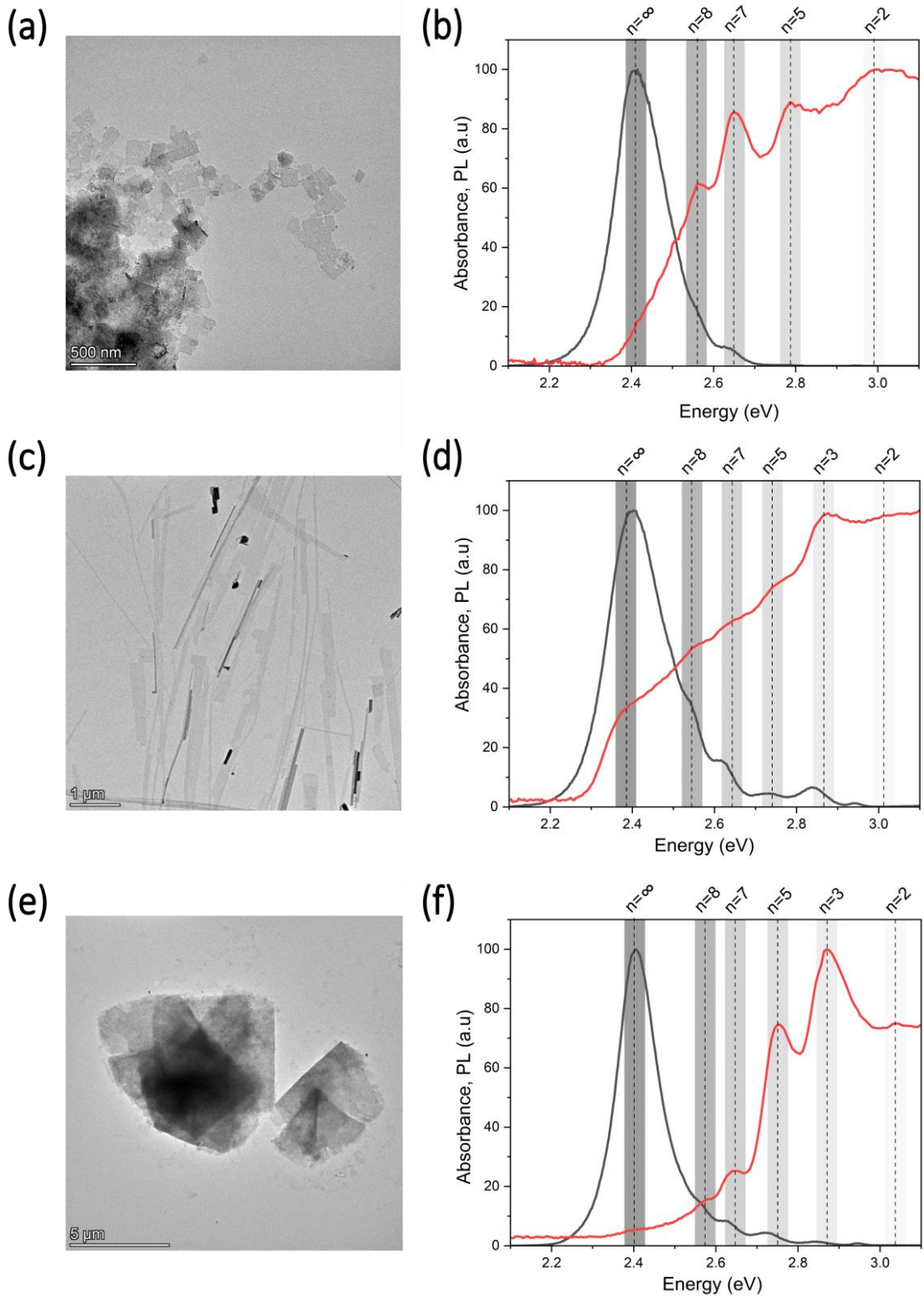

**Figure 1.** TEM images with corresponding absorbance and PL spectra of quasi-2D MAPbBr$_3$ nanoparticles with different morphologies (a, b) nanoplatelets (c, d) nanostripes and (e, f) nanosheets. The spectral features, marked by vertical bars in the absorbance and PL spectra, represent different n-phases.

**Photodetection**

The photodetection process is governed by (i) charge carrier generation by illumination with light and (ii) transport and collection of carriers at the electrodes. The photodetectors produce charge carriers as excitons (electron-hole pairs), free electrons and holes. Strongly bound excitons are more predominant in low dimensional perovskites, whereas weakly bound or free charge carriers in bulk perovskites due to their exciton binding energies.[12] The bulk perovskites have exciton binding energies comparable or less than thermal energy at room temperature resulting in faster (several picoseconds) exciton dissociation to free electrons. This non-excitonic nature leads to effective charge generation and collection compared to low dimensional perovskites having much larger exciton binding energies (tens to hundreds meV) and stronger confinement effect.[26] In quasi-2D perovskites, the extent of confinement and binding energy are tunable which plays a vital role in photodetection. The quasi-2D perovskites in weaker confinement (high n-phases) have lower exciton binding energy favoring quicker exciton dissociation at ambient temperatures.[26,27] In mixed-phase quasi-2D perovskite, the distribution of different phases can be engineered for directional and effective carrier transport. Additionally, effective photodetection requires high carrier mobility, long diffusion length, long lifetime, low recombination rate, and low trap densities.

To study the photodetection performance of the three different morphologies of quasi-2D perovskites, the lateral configuration devices (Au/perovskite/Au) were fabricated by depositing the nanoparticles on gold electrodes as shown in the inset of Figure 2 (a) and S5. The deposited particles form a network to bridge the electrode channel length (Figure S6). A single conductive path that a carrier has to pass in nanoplatelets consists of many particles due to their smaller size. This leads to a hopping charge transfer between neighboring nanoplatelets. Nanostripes and nanosheets due to their larger size (> channel length 5 μm) can bridge the channel as shown in Figure S6 (b, c) and provide a direct charge transport pathway. However, not all the nanostripes or nanosheets contribute to making a direct pathway due to random distribution and polydispersity in their shape and size as shown in Figure S6 (b, c).

The deposited nanocrystals were annealed in a vacuum. After annealing, the photocurrent increased by a factor of three with a reduction in dark current by a factor of three as shown in Figure S7. The annealing process results in the ligand's partial removal, leading to better contact with electrodes and free charge carriers' transport.[27]

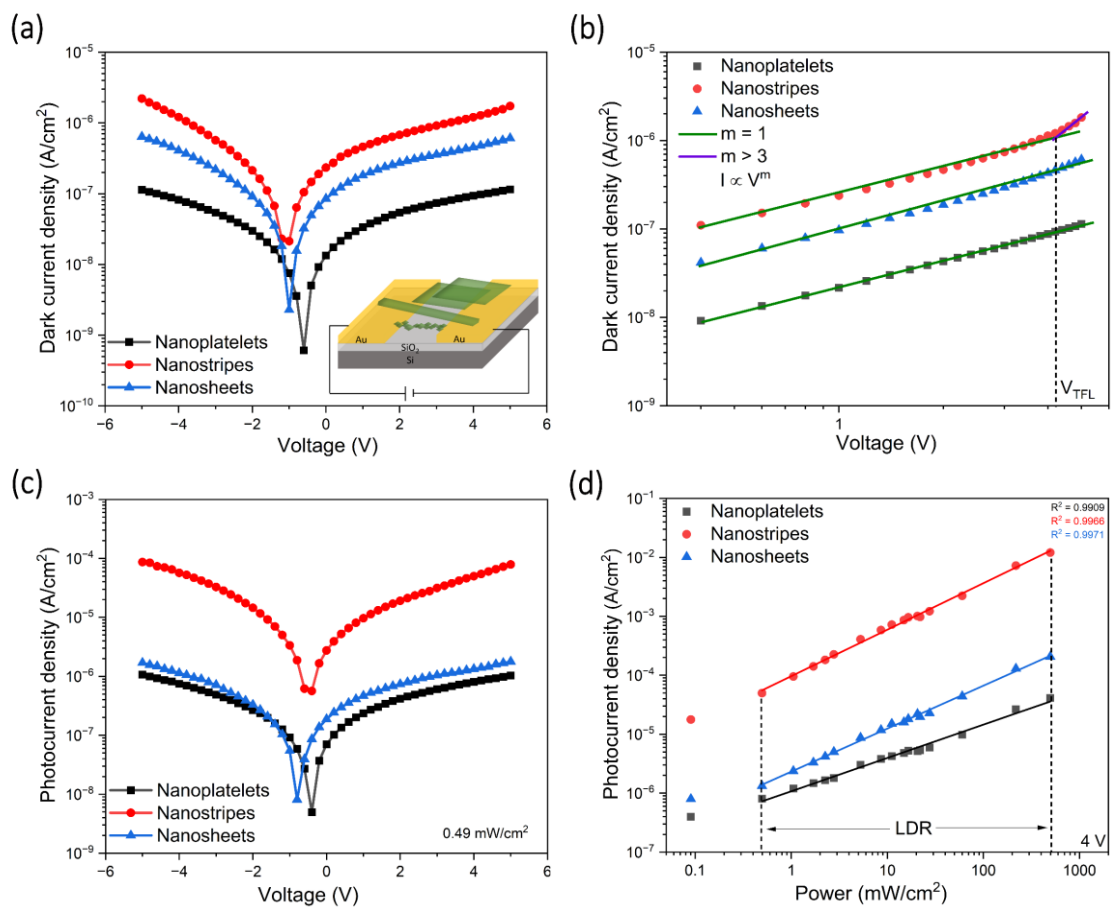

**Figure 2.** Current-voltage (I-V) measurements in the range -5 to +5 V. (a) dark current density, inset shows a schematic of quasi-2D MAPbBr$_3$ with different morphologies deposited on the gold electrodes (b) dark current density of nanoplatelets and nanosheets show linear ohmic regime ($I \propto V$, green line). Nanostripes show a linear ohmic regime followed by a trap-filled regime ($I \propto V^3$, purple line) marked by a steep increase in current at higher voltages. (c) photocurrent density measured for different morphologies of MAPbBr$_3$ by illuminating laser ($\lambda_{ex}$ = 405 nm) with power density 0.49 mW/cm$^2$ (d) photocurrent measured at different illumination power at a bias voltage of 4 V displaying wide linear dynamic range (LDR).

The current-voltage (I-V) measurements for three different morphologies of quasi-2D colloidal perovskites under darkness and light illumination are shown in Figure 2. The dark current for different morphologies is compared in Figure 2 (a). A dark current of $5.4 \times 10^{-8}$ A/cm$^2$ for nanoplatelets, $6.8 \times 10^{-7}$ A/cm$^2$ for nanostripes, and $2.7 \times 10^{-7}$ A/cm$^2$ for nanosheets are observed at a bias voltage of 2 V. The lower dark current in nanoplatelets and nanosheets indicates the presence of a low density of intrinsic free charge carriers due to the layered structure (with insulating alkyl-long chain) and the stronger confinement effect.[28] In addition to the confinement effect, the hopping transfer of carriers in nanoplatelets results in the suppressed dark current. A significant hysteresis between forward and backward scans with zero-point drifts is observed as shown in Figure S8. Perovskites inherently have high ionic mobility along with the electronic contribution to the transport and it has been reported that the movement of ions causes hysteresis in the I-V characteristics.[29,30]

The space charge limited current (SCLC) analysis is shown in the logarithmic I-V response in Figure 2 (b) for different morphologies. An Ohmic region with linear response ($I \propto V$, green line) is observed at a lower voltage for all three morphologies. At higher voltages for nanostripes a nonlinear steep increase in current ($I \propto V^3$, purple line) is observed. This reveals a transition from Ohmic to trap filling limit (TFL) region at $V_{TFL}$= 4.2 V. At the TFL transition, all the trap states are occupied by charge carriers and the trap density $n_t$ can be estimated by [31]

$$n_t = \frac{2\varepsilon_r \varepsilon_0 V_{TFL}}{eL^2}$$

where $\varepsilon_r$ is the relative dielectric constant (25.5), $\varepsilon_0$ is the vacuum permittivity, $e$ is the elementary charge and $L$ is the channel length (5 µm). The obtained trap density for nanostripes is $4.7 \times 10^{14}$ cm$^{-3}$. The transition from the Ohmic to the TFL region was not observed in the measurement range (-5 to +5 V) for nanoplatelets and nanosheets and V$_{TFL}$ will occur beyond 5 V. This will result in a higher density of traps for the nanoplatelets and nanosheets compared to nanostripes. The photocurrent measured at illuminating light of power density 0.49 mW/cm$^2$ is shown in Figure 2 (c). The nanoplatelets and nanosheets exhibit a similar order of photocurrent. The nanoplatelets show a photocurrent of $4.2 \times 10^{-7}$ A/cm$^2$, which is 7.7 times its dark current. The nanosheets show a photocurrent of $7.3 \times 10^{-7}$ A/cm$^2$, 2.7 times its dark current. The nanostripes exhibit the highest photocurrent of $1.9 \times 10^{-5}$ A/cm$^2$, 28 times its dark current. The better photocurrent in nanostripes is due to effective carrier generation, exciton dissociation, lower trap density and transport through the network of the majority of high-n phases. Figure 2 (d) shows logarithmic photocurrent with different illumination power densities for different morphologies. The photocurrent linearly increases with an increase in laser power density for all morphologies. The increase in photocurrent is due to the photo-induced generation of carriers (electron-hole pairs) at higher power densities. It reveals a wide linear dynamic range (LDR) given by

$$LDR = 20 \log \left( \frac{I_{max}}{I_{min}} \right)$$

where $I_{max}$ and $I_{min}$ are the maximum and minimum photocurrent of the linear range, respectively. The obtained LDR for nanoplatelets, nanostripes, and nanosheets are 34.2 dB, 47.7 dB, and 43.8 dB, respectively. The I-V response with increasing power densities for all the morphologies is shown in Figure S9. It can be noted that the shape of the I-V curves is nonlinear with the extent of nonlinearity being maximum in nanostripes and minimum in nanoplatelets. The nonlinear behavior indicates the formation of a Schottky barrier between the gold electrode and perovskite.[32] The fitting with the power law $I = AP^\theta$, where A is a constant and $\theta$ is the exponent which defines the response of the current to the light power. It gives $\theta$ = 0.57, 0.78, and 0.71 for nanoplatelets, nanostripes, and nanosheets respectively as shown in Figure S9 (d). As the value is 0.5<$\theta$<1, it indicates the presence of defects and traps in the perovskites leading to complex electron-hole generation and recombination processes as reported earlier. [33]

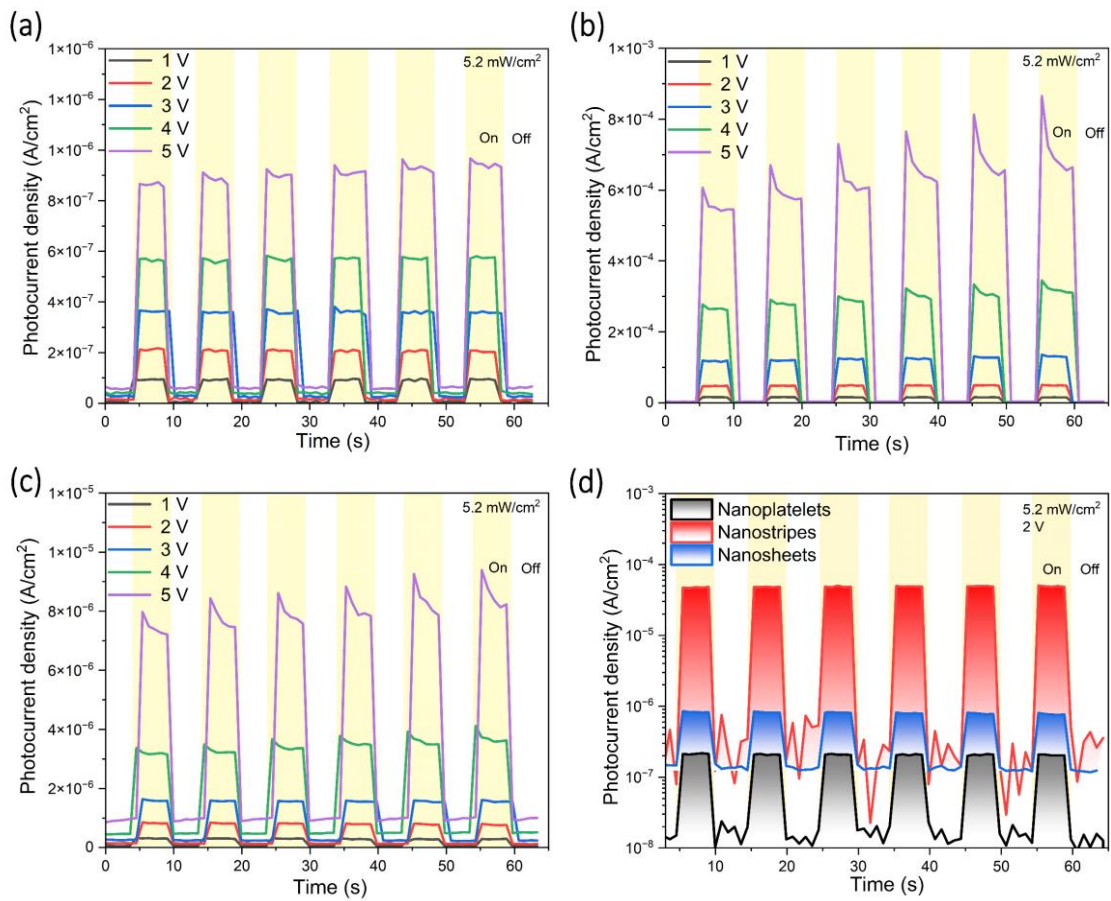

**Figure 3.** Transient photocurrent density measured at 5.2 mW/cm$^{-2}$ illumination power for different bias voltages using an optical signal with a period of 10 s (a) nanoplatelets (b) nanostripes and (c) nanosheets of MAPbBr$_3$ (d) Transient photocurrent density measured at fixed bias voltage 2 V and illumination power 5.2 mW/cm$^{-2}$ for different morphologies of MAPbBr$_3$.

We measured the transient photocurrent density at various bias voltages as shown in Figure 3 (a-c) for different morphologies by periodically switching the laser (5.2 mW/cm$^{-2}$) on/off for the duration of 10 s. The photocurrent increases with the bias voltage. A large number of electron-hole pairs are generated with higher bias voltage accounting for a rise in photocurrent. Similarly, an increase in laser power density results in enhanced photocurrent, as shown in the transient photocurrent measured at various illumination powers at the constant bias of 2 V (Figure S10). The overshooting spike followed by photocurrent decay is observed for all the morphologies above the voltage of 4 V. The magnitude of a spike is higher in the nanostripes and nanosheets compared to nanoplatelets. The spike in the transient response is a widely observed feature in perovskites. [34] This is attributed to capacitive charging due to mixed electronic-ionic nature which leads to electronic or ionic charge accumulation at the electrode. The mobility of ions and charges in the nanoplatelets are suppressed by hopping transport between the nanoplatelets substantially increasing their resistance and thus the charging time constant ($\tau = RC$). If the charging time is small compared to duration of the pulse, the capacitor is charged slowly which is observed in the current transients as a stable signal without a spike. The transient photocurrent density for different morphologies is compared in Figure 3 (d). The highest on/off ratio is observed in nanostripes followed by nanoplatelets and nanosheets. The better on/off ratio in nanoplatelets compared to nanosheets is attributed among other factors to a higher PLQY which compensates for the limitations of hopping carrier transport. The higher PLQY in nanoplatelets can be considered as a consequence of reduced non-radiative rates in nanocrystals. This in turn means that the total recombination losses are lowered leading to higher quantum efficiency of the photodetector.[35] Additionally, the dark current in nanoplatelets is smaller than in nanosheets obviously due to the substantially larger number of potential barriers across the film of nanoplatelets between individual nanocrystals (mentioned higher resistivity).

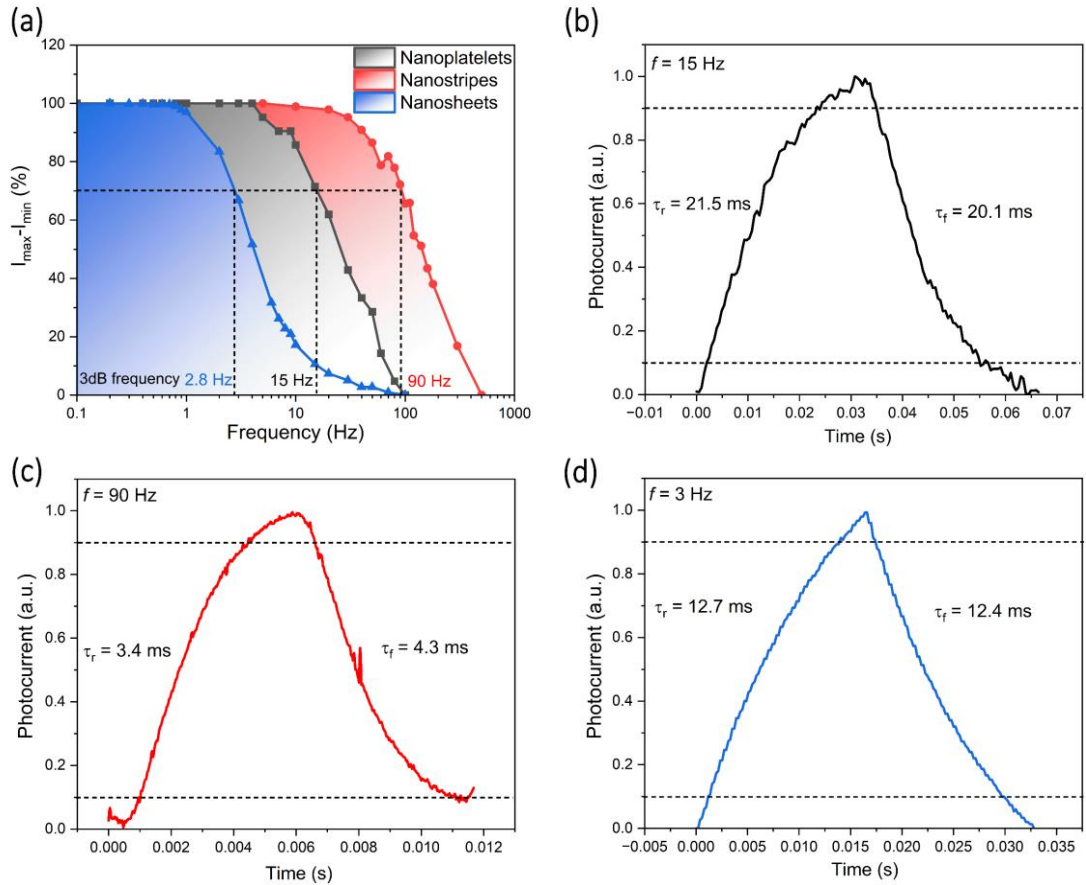

**Figure 4.** Photoresponse characteristics at 4 V bias and pulse light illumination power 5.2 mW cm$^{-2}$ at different frequencies (a) temporal response and 3dB bandwidth. Estimating rise time ($\tau_r$) and fall time ($\tau_f$) with rising and falling edges for different morphologies of MAPbBr$_3$ (b) nanoplatelets (c) nanostripes and (d) nanosheets.

The frequency response was recorded by pulsing the laser at different frequencies from 0.1 to 150 Hz and noting the normalized relative photocurrent (I$_{max}$-I$_{min}$) for all the morphologies as shown in Figure 4 (a). The 3dB frequency or bandwidth indicates the frequency at which the photodetector's response falls to (I$_{max}$-I$_{min}$)/√2 (or 70.7% of its maximum value). This is a crucial parameter for developing photodetector-based high-speed optical communication systems, which demand the capability to manage faster data rates. The nanostripes exhibit the highest 3dB frequency of 90 Hz. The nanoplatelets and nanosheets show a 3dB frequency of 15 Hz and 2.8 Hz respectively. The nanostripe-based photodetectors can operate at much higher frequencies than nanoplatelets and nanosheets. Figure 4 (b, c, d) shows the response time evaluated at 3dB frequency for nanoplatelets, nanostripes, and nanosheets. The rise time ($\tau_r$) and fall time ($\tau_f$) obtained for the nanoplatelets are 21.5 ms and 20.1 ms respectively. A faster response time $\tau_r$=12.7 ms and $\tau_f$=12.4 ms is observed for nanosheets. The nanostripes display the fastest response time with $\tau_r$=3.4 ms and $\tau_f$=4.3 ms compared to nanoplatelets and nanosheets. This is attributed to less pronounced charge carriers trapping and rapid transfer because of higher mobility inside a bulk-like domain with shorter lifetime.[36] Additionally, the fast response is also attributed to nanostripes' morphology which well connects the electrode spacing. These response times surpass many previously reported values for perovskite nanoparticle-based photodetectors as listed in Table S3.

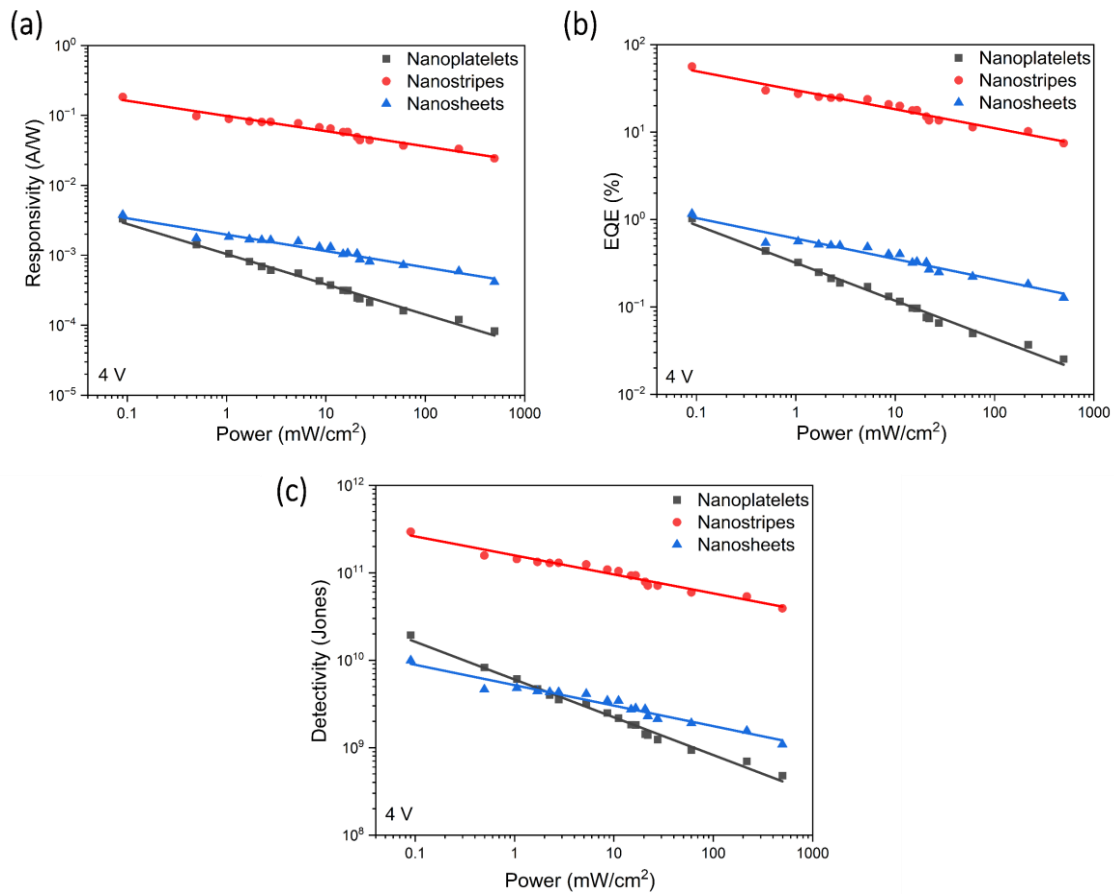

**Figure 5.** Photodetection performance (a) responsivity (b) EQE and (c) detectivity for different morphologies of MAPbBr$_3$ at 4 V bias.

To further compare the photodetection performance of different morphologies of perovskite, photodetector parameters, responsivity (R), external quantum efficiency (EQE), and detectivity (D) were evaluated using the expressions. [18]

$$R = \frac{I_{photo} - I_{dark}}{PA}$$

$$EQE = \frac{Rhc}{e\lambda}$$

$$D = \frac{R}{\sqrt{2e(I_{dark}/A)}}$$

where $I_{photo}$ and $I_{dark}$ are photocurrent and dark current respectively. $P$ is the illumination light power density, $A$ is the active area of the photodetector, $h$ is Planck constant, $c$ is the speed of light, $e$ is the elementary charge and $\lambda$ is the wavelength of the laser.

The photodetection parameters with increasing illumination power density at the bias of 4 V for all the morphologies of perovskite are shown in Figure 5. The highest values of R, EQE and D are observed at the low power density of 0.09 mW cm$^{-2}$. It declines with further increase ascribing to a predominance of bimolecular recombination at higher power densities, which agrees with earlier publications.[37] The nanoplatelets and nanosheets show almost identical values of R and EQE at lower power density, with 3.4 mA W$^{-1}$ and 1.0 % for nanoplatelets and 3.8 mA W$^{-1}$ and 1.2 % for nanosheets. However, at higher power densities the nanosheets exhibit better values of R and EQE compared to the nanoplatelets. The better performance in nanosheets compared to nanoplatelets is attributed to the effective bridging of electrodes due to their lateral size and reduced carrier hopping and interfacial scattering. The nanoplatelets and nanosheets exhibit detectivities of 1.9 x 10$^{10}$ Jones and 9.9 x 10$^9$ Jones respectively. At low power density, the detectivity values for nanoplatelets are higher than for nanosheets due to their ultra-low dark current. But as power density exceeds 1.7 mW cm$^{-2}$ the nanoplatelets show reduced values compared to nanosheets. This indicates the degradation of nanoplatelets on facing higher illumination power along with the increased resistance due to carrier hopping.[38] The nanostripes exhibit the highest R, EQE, and D of 183 mA W$^{-1}$, 56 % and 2.9 × 10$^{11}$ Jones compared to nanoplatelets and nanosheets. The R, EQE, and D with various bias voltages at a constant power density of 0.09 mW cm$^{-2}$ are shown in Figure S11 for all three morphologies. The R, EQE and D values increase with voltage.

Many studies have reported photodetectors based on colloidal or nanoparticle-based perovskites such as nanowires, nanosheets, etc. as shown in Table S3. Our nanostripes exhibit high performance among MAPbBr$_3$-based photodetectors, which can be further enhanced by fabricating nanoscale photodetectors by contacting individual nanorod as demonstrated in an earlier publication.[39]

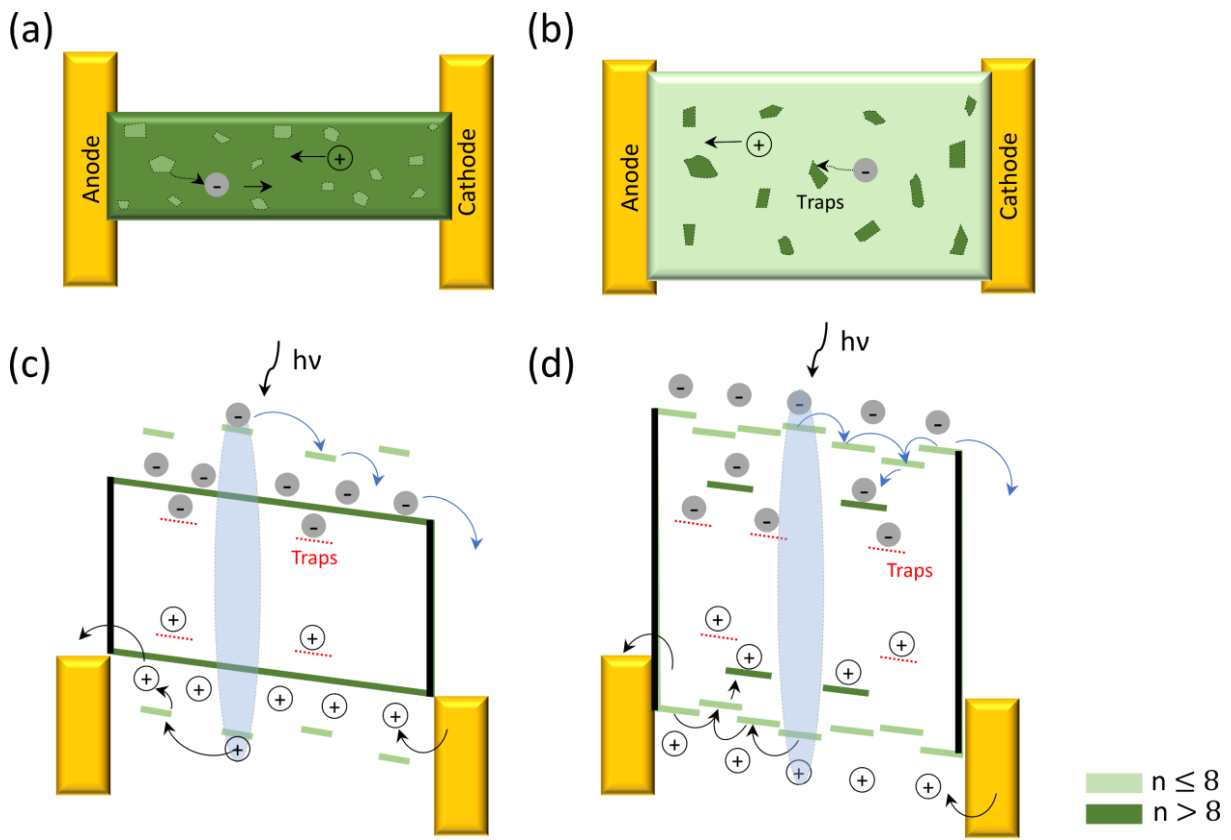

**Figure 6.** Schematic illustration of charge carrier transport and separation in (a, c) nanostripe and (b, d) nanosheet-based quasi-2D MAPbBr$_3$ photodetector. The high-n phases (n>8) are indicated by dark green regions and low-n phases (n≤8) are indicated by light green regions.

The better photodetection performance by the nanostripes can be understood from the charge carrier transport mechanism schematics shown in Figure 6. The gold electrode collects and distributes electrons and holes through quasi-2D MAPbBr$_3$ nanoparticles. With gold electrodes, having a work function of -5.1 eV, the photoexcited hole's movement is much easier creating a hole-dominated device. The individual nanostructures are consisting of mixed-n phases, the carrier transport is significantly associated with the n values. The nanostripes primarily consist of high-n phases (bulk-like regions with n>8 are shown in dark green in Figure 6 (a)), while the nanosheets are predominantly composed of low-n phases (highly confined regions with n≤8 are shown in light green in Figure 6 (b)). The dominant bulk-like phase in nanostripes provides an effective network for carrier transport. The quantum and dielectric confinement effect weakens in nanostripes, leading to enhanced carrier generation by faster exciton dissociation to free electrons. In addition, it extracts carriers from the small fractions of low-n phase regions through relaxation to the band-edge as shown in Figure 6 (c). The high-n phases suppress the minority low-n phases in nanostripes, resulting in longer diffusion lengths.[40] In nanosheets, dominant low-n phases make the exciton dissociation into free charge carriers challenging due to a stronger quantum and dielectric confinement effect. Additionally, spatially isolated small fractions of high-n phases act as effective funnels and eventually collect and trap charge carriers hampering the responsivity, detectivity, and response time. Playing an important positive role for the photoluminescence, funnelling effect can compete with trapping of excitons giving rise to high QY and longer PL lifetimes. However, in the mixed-n perovskites it leads inevitably to the trapping of charge carriers in the funnels itself impeding the electrical transport. The carrier transport mechanism in nanoplatelets is similar to the nanosheets with additional hindrance emerging from carrier transferring between particles. Despite having a shorter lifetime and lower PLQY than nanosheets, nanostripes exhibit superior electrical transport and photodetection performance. This is obviously due to the optimal composition of high-n and low-n phases. As a result, the structures exhibit the effective drifting of generated carriers to the electrodes through a network of high-n phases and effective extraction of exciton through dissociation and funneling to weaker confinement regions.

## Conclusion

We synthesized three different morphologies (nanoplatelets, nanostripes, and nanosheets) of colloidal quasi-2D $MAPbBr_3$ perovskite using the hot injection method with slight variation of reaction parameters and investigated their photodetection performance. The absorbance and PL spectra demonstrate the presence of mixed-n phases leading to different extents of quantum confinement. The nanostripes have the weakest and nanosheets have the strongest confinement effect. Nanoplatelets and nanosheets exhibit higher PLQY and longer lifetimes than nanostripes. The nanostripes exhibit the highest photocurrent of $1.9 \times 10^{-5}$ $A/cm^2$ and highest on/off ratio and the fastest photoresponse ($\tau_r$=3.4 ms and $\tau_f$=4.3 ms) compared to nanoplatelets and nanosheets. Nanostripes show better photoresponsivity, EQE and detectivity with values of 183 mA $W^{-1}$, 56 % and $2.9 \times 10^{11}$ Jones, respectively, outperforming nanoplatelets and nanosheets. The superior photodetection by the nanostripes is attributed to two key factors. First, a weaker confinement effect leads to faster and more effective exciton dissociation to free charge carriers. Second, the presence of a high-n phase network facilitates effective carrier transport across the photosensitive nanostructure. This study reveals that effective charge carrier transport can be achieved by strategically engineering the distribution and composition of high-n phases in nanoparticles. Our investigation will contribute to the fundamental understanding required to develop a high-performance photodetector based on mixed-n quasi-2D perovskites.

# Supplementary Information

The absorbance and PL studies of MaPbBr$_3$ nanoplatelets, nanostripes and nanosheets. SEM, optical and fluorescence microscope images of different morphologies. Current-voltage (I-V) and transient photocurrent measurements at different illumination power for nanoplatelets, nanostripes and nanosheets. Photodetection performance for different morphologies. Comparison of photodetection performance of nanoparticle-based perovskites. (PDF)

# Author Contributions

The manuscript was written through contributions of all authors. All authors have given approval to the final version of the manuscript.

# Conflicts of interest

There are no conflicts to declare.

# Acknowledgments

B.M.S. acknowledges Alexander von Humboldt-Stiftung/Foundation for the postdoctoral research fellowship. B.M.S also acknowledges Ronja Piehler for insightful discussions and technical help. Deutsche Forschungsgemeinschaft (DFG, German Research Foundation) is acknowledged for funding of SFB 1477 "Light-Matter Interactions at Interfaces", project number 441234705, W03 and W05. C. K. also acknowledges the European Regional Development Fund of the European Union for funding the PL spectrometer (GHS-20-0035/P000376218) and X-ray diffractometer (GHS-20-0036/P000379642) and the DFG for funding an electron microscope Jeol NeoARM TEM (INST 264/161-1 FUGG) and an electron microscope Thermo Fisher Talos L120C (INST 264/188-1 FUGG).

# Abbreviations

RP, Ruddlesden-Popper; TOP, tri-octylphosphine; BTD bromotetradecan; MAB, methylammonium bromide; DPE, diphenyl ether; DDA, dodecylamine; PL, photoluminescence; TRPL, time-resolved photoluminescence; PLQY, Photoluminescence quantum yield; SEM, scanning electron microscope; SCLC, space charge limited current; TFL, trap filling limit; LDR, linear dynamic range; R, responsivity; EQE, external quantum efficiency; D, detectivity.